\documentclass{aastex631}
\usepackage{multirow}
\usepackage{gensymb}
\usepackage{xcolor}
\usepackage{amsmath}
\usepackage{xcolor}

\begin{document}
\title{Fallback rates in partial tidal disruptions of white dwarfs by intermediate mass black holes}
\correspondingauthor{Tapobrata Sarkar}
\email{tapo@iitk.ac.in}

\author{Debojyoti Garain}
\affiliation{Department of Physics, Indian Institute of Technology Kanpur,
	Kanpur 208016, India}

\author{Tapobrata Sarkar}
\affiliation{Department of Physics, Indian Institute of Technology Kanpur,
	Kanpur 208016, India}

\begin{abstract}
Fallback rate of debris after a partial tidal disruption event of a star with an intermediate mass black hole (IMBH) might provide important signatures
of such black holes, compared to supermassive ones. 
Here using smoothed particle hydrodynamics methods, we provide a comprehensive numerical analysis of this phenomenon. We perform 
numerical simulations of single partial tidal disruptions of solar mass white dwarfs in parabolic orbits, with a non-spinning $10^3M_{\odot}$ 
IMBH for various values of the impact parameter, and determine the core mass fractions and fallback rates of debris into the IMBH. 
For supermassive black holes, in a full disruption processes, it is known that the late time fallback rate follows a power law $t^{-5/3}$, whereas for 
partial disruptions, such a rate has been recently conjectured to saturate to a steeper power law $t^{-9/4}$, independent of the mass 
of the remnant core. We show here that for IMBHs, partial disruptions significantly alter this conclusion. That is, 
the fallback rate at late times do not asymptote to a $t^{-9/4}$ power law, and this rate is also a strong function of the core mass. We derive a robust formula 
for the late time fallback rate as a function of the core mass fraction, that is independent of the white dwarf mass, as we verify numerically by varying the mass 
of the white dwarf. 
\end{abstract}
	
\section{Introduction}
\label{sec1}

Intermediate mass black holes (IMBHs, see \cite{Green} for a recent review) have been dubbed as the `missing link' in the 
theory of black hole (BH) formation and evolution. These BHs have masses $\sim 10^2 - 10^5 M_{\odot}$ and are believed to exist, for example, in
dwarf galaxies and globular clusters, and are possible seeds of supermassive black holes (SMBHs), see e.g., \cite{Volonteri}. 
There are however only a handful of recent observational evidences for these \citep{IMBH0, 
IMBH1, IMBH2, IMBH3, IMBH4}, in contrast to the many confirmed observations of SMBHs and stellar mass black holes. 
Physical properties of IMBHs are thus important and interesting to study further. An important phenomenon that can serve as a distinctive
feature of IMBHs are tidal disruption events (TDEs), since electromagnetic emission arising out of these events have a strong 
dependence on the BH mass. Generically, TDEs from BHs produce a luminous flare, and several such events have
been detected \citep{Holoien}. Light curves from TDEs have been shown to broadly conform to theoretical predictions 
\citep{Brown}, and hence, observations of TDEs from IMBHs might be a crucial signature of the latter. In this context, 
another well studied object in stellar astrophysics is the tidal disruption of white dwarf (WD). The sufficiently accurate equation of state of
a WD makes it an attractive object to study in this scenario. Now, it is well known that WDs can only be tidally disrupted by IMBHs and 
are captured by SMBHs, see the book chapter by Maguire et. al. in \cite{Jonker}. TDEs involving WDs and IMBHs can thus
be an ideal laboratory to glean further insight into the distinction between IMBHs and SMBHs. With this motivation, the purpose of
this paper is to perform a comprehensive analysis of partial TDEs involving WDs in the background of IMBHs. 

Recall that a stellar object is tidally disrupted by a BH when its self gravity is overcome by the non-local gravitational effects of the BH. 
From a Newtonian perspective, a TDE occurs when the pericenter position $r_p$ of a spherical star of 
mass $M_{\star}$ and radius $R_{\star}$ comes within the tidal radius $r_t\sim R_{\star}(M_{\rm BH}/M_{\star})^{1/3}$, with $M_{\rm BH}$ being 
the mass of the BH, see \cite{Hills} and related early works by \cite{FrankRees}, \cite{Lacy1}, \cite{CarterLumineta}, \cite{CarterLuminetb}. 
This is obtained by simply equating the tidal force on the spherical stellar object to its self gravity,
but  is often used as an order of magnitude estimate in more realistic situations when a star can be considerably deformed 
before being disrupted. In the presence of a tidal force, a star can be fully disrupted wherein it breaks up completely and is reduced to 
a stream of debris \citep{Kochanek1994, Guillochon2014}, or the disruption can be partial, leading to a remnant core, along with the debris \citep{Manukian2013, Gafton2015, Banerjee2023}. Once full tidal disruption has taken place, 
roughly half of the debris is bound to the BH and in time returns to the pericenter and falls into the BH. The situation is modified in
partial disruption processes, but in due course, for both full and partial disruptions, the in-falling 
material dissipates energy and results in accretion flow, and can cause a luminous event \citep{Evans1989, Hayasaki2013, Hayasaki2016, 
Bonnerot2016, Liptai2019, Clerici2020}. Also, the process of partial tidal disruption can in principle occur multiple times \citep{MacLeod2013}, 
if initial conditions allow the remnant core to come back to a new pericenter position. TDEs have been extensively investigated
over more than four decades now, and continue to be an active research area in stellar astrophysics. 
For a sampling of the more recent literature, see, e.g., \cite{Lodato}, \cite{Guillochon2013}, \cite{Rosswog_TR}, \cite{Coughlin2015}. 
Other studies include the effect on TDEs due to the black hole's mass 
and spin \citep{Chen2018, Gafton2019, Ryu2020a, Wang2021}, stellar structure \citep{Golightly2019a, Law2019, Ryu2020b}, stellar rotation \citep{Golightly2019b, Kagaya2019, Sacchi2019, Law2017} and stellar orbital parameters \citep{Darbha2019, Milesetal, Park2020, 
Cufari2022}, to name a few. A comprehensive recent review of various features of TDEs can be found in \cite{Jonker}. An updated catalog of candidate TDEs can be 
found at the \cite{OpenTDE}.

Ubiquitous in the study of tidal disruptions -- both full and partial -- is the notion of the fallback rate of bound debris into the BH. 
This is an important quantity to study, since it influences several features of light curves from TDEs. The fallback of
debris starts when the most bound debris returns to the BH, with the rate of fallback reaching a maximum (called the peak fallback rate), and then 
asymptotes to a power law behaviour with time, for late times. A fundamental quantity of interest in this context is this late-time power
law scaling, i.e., the mass fallback rate of tidal debris ${\dot M} \sim t^{n}$. A considerable amount of literature exists on the
topic till date, with the initial result by \cite{Rees} that $n = -5/3$ (the work \cite{Rees} quotes the exponent as $-5/2$ and this was 
subsequently corrected by \cite{Phinney} to $-5/3$, see also \cite{Ulmer1999}). The argument is simple and only depends on 
Kepler's third law. Although the process of
tidal disruptions is reasonably complicated and the forces on the debris are not strictly central even at late times due to the
presence of gravity among the debris particles, it is still a good estimate at late times, as shown by \cite{Lodato}. 
This is because at late times, the density of the remaining debris is small, and Keplerian dynamics is often a reasonable approximation. While this power law
has been tested for full TDEs \textcolor{black}{in a number of notable works, see, e.g., \cite{Evans1989}, \cite{Coughlin2015}, \cite{Law2020}}, 
it was established more recently that partial disruptions might 
considerably alter the situation. The essential point here is that after a partial disruption process, 
the core exerts gravitational influence on the debris that is bound to the BH, a fact that is irrelevant in full TDEs where 
there is no core to begin with. Now, the late time fallback rate 
is considerably influenced by the core, since particles falling back into the BH at such times originate close to the core itself. 
Naively, this is associated with the Hill radius of the core -- a distance up to which gravitational influence of the core is
greater than that of the BH. 

The first indication that the late time peak fallback rate might be steeper than the $t^{-5/3}$ power law was provided
by \cite{Guillochon2013} where it was observed that the exponent was, in some cases, close to $-2.2$ for TDEs by
SMBHs. Following this, recently, \cite{CoughlinNixon} provided an analytical treatment 
of the power law behaviour of the late time peak fallback
rate from SMBHs. They found that for full TDEs from such BHs, ${\dot M} \sim t^{-5/3}$ 
in accordance with previous analytical results and simulations, but that for partial TDEs, ${\dot M} \sim t^{-9/4}$. 
The latter represents a significant departure from the full TDE case, and it was argued by  \cite{CoughlinNixon} that
this power law is effectively independent of the core mass fraction. As pointed out by \cite{Ryu2020c}, this analysis has a number of 
assumptions, one of the main ones being that the force on the debris by the core is radial in direction. 
Indeed, the dynamics of the debris even at late times is complicated especially in the presence of a remnant core. 
Taking the Hill radius as a rough estimate of the influence of the core, one finds (both analytically and numerically) that this radius increases with time,
and hence the core should continue to influence the falling debris even at late times. This indicates that Keplerian
dynamics would be strictly invalid even at such late times. The behaviour of the late time slope of the fallback rate
should therefore be carefully checked with numerical simulations. 
This was indeed done by \cite{Milesetal}, who found from simulations of a $1 M_{\odot}$ polytropic star in the
background of a non-rotating $10^6 M_{\odot}$ BH that ${\dot M} \sim t^{-9/4}$ was 
reasonably accurate and that the exponent was effectively independent of the remnant core, as suggested by analytical results. 
However, immediately afterwards, \cite{Wang2021} performed a similar analysis of TDEs involving stellar mass black holes. They find
that the late time fallback rate was significantly different from a $t^{-9/4}$ law. 

In this paper, we ask the question if the same is true for the TDE of a $1 M_{\odot}$ WD by an IMBH. 
The main motivation for this study is that, as mentioned in the beginning of this paper, there is currently 
not enough compelling observational evidence of IMBHs, 
and in such a situation, any further analysis that can shed light on the physical properties of IMBHs should be important and interesting. 
WDs that come close to such an IMBH will be tidally disrupted either fully or
partially, and such TDEs can play a vital role in detecting these objects. Further, accurate equations of state for WDs are
well established, and using these, one can effectively revisit the results of \cite{CoughlinNixon} to see to what extent 
the late time slope of the fallback rate, originally derived for SMBHs are applicable to IMBHs. Indeed, as we show in the course of this paper, we find
significant departures from the results of \cite{CoughlinNixon}. One of our main results is that for IMBHs, there is a strong dependence
of the late time slope of the fallback rate on the remnant core mass, contrary to the SMBH case. Here, we consider single-encounter, full and
partial disruptions of WDs in parabolic orbits. A recent work by \cite{Chen2023} considers multiple tidal interactions of a WD that is
in-spiralling into an IMBH. 

The purpose of the rest of the paper then is to carry out a comprehensive analysis of TDEs involving a solar mass WD
and a $10^3M_{\odot}$ IMBH. For various values of the impact parameter. This is done using using 15 smoothed particle hydrodynamics simulations
of such events by varying the impact parameter.  We study (a) the core mass fraction, (b) the peak fallback rate, 
(c) the instantaneous power law index of the fallback rate and (d) the late time slope of the fallback rate as functions
of the impact parameter. In addition, 7 more numerical simulations were performed to check the robustness of our results
for WDs with different masses, while keeping the black hole mass fixed. In the next section \ref{sec2}, we describe the methodology, 
followed by the results of our numerical analysis in section \ref{sec3}. Section \ref{sec4} concludes this paper with a summary of our main results. 

\section{Methodology}
\label{sec2}

In this section, we present our numerical methodology for simulating the tidal disruption of WDs by an IMBH. For a detailed description of the numerical methods employed
here, we refer the reader to \cite{Banerjee2023} and avoid repeating the details here for brevity. Our simulation of the fluid star uses the smoothed particle hydrodynamics (SPH) technique, which discretizes the fluid star into a set of particles, each possessing density, position, velocity, and other properties. To efficiently compute fluid properties and forces on each particle, we implement a binary tree with a tree opening angle parameter set at $\theta = 0.5$. Standard SPH artificial viscosity parameters, $\alpha^{\rm {AV}} = 1.0$ and $\beta^{\rm {AV}} = 2.0$ (see \cite{Hayasaki2013}), are adopted, while the Balsara switch is used to reduce the viscosity in shear flows (see \cite{Balsara1995}). The SPH fluid equations are integrated using the leapfrog integrator, ensuring the exact conservation of energy and angular momentum throughout the simulation. In modeling the external gravitational influence on each particle originating due to the Schwarzschild black hole, we adopt the same approach as detailed in \cite{Banerjee2023}. In this work, we employ a global time step for the evolution of the system.

We perform two-stage numerical simulations to study tidal disruption events. Initially, we create a spherically symmetric WD in equilibrium. Once the WD is modeled, we induce tidal disruption by setting the WD in motion under the gravitational influence of the black hole. 
To attain the equilibrium structure of a carbon-oxygen WD using the SPH formalism, we follow the methodology described in \cite{Garain2023}. Hydrostatic equilibrium within the WD is attained when the pressure resulting from degenerate electrons balances the self-gravitational forces causing its collapse. The pressure due to degenerate electrons and the density originating from carbon atoms within the WD are expressed as follows:

\begin{eqnarray}
		P  = K_{P} \Bigl[x(1+x^2)^{1/2}(2x^2/3-1)+ \log_{e}\bigl[x+(1+x^2)^{1/2}\bigr] \Bigr], ~
		\rho  =  K_{\rho} x^3
\end{eqnarray}

where the constants are defined as $K_{P} = 1.4218\times10^{24}/(8\pi^2)$  $\rm N~m^{-2}$, $K_{\rho} = 1.9479 \times 10^9$ $\rm kg~m^{-3}$, and $x$ represents the `relativity parameter' (dimensionless Fermi momentum). Consequently, the pressure and density are related through the parameter $x$, which serves as the equation of state (EOS) for our study. In SPH, particles are initially distributed in a closed-packed sphere, which is then stretched to attain the density profile for the WD, which is obtained by employing the methodology outlined in \cite{Garain2023}. Subsequently, this stretched SPH profile of the fluid star is evolved in isolation, without the presence of the black hole, in order to smoothen out the density fluctuations. The final relaxed profile of the fluid star is achieved once its density profile matches exactly with the WD density profile. 

In our study, we consider an IMBH with a mass $M_{\rm BH} = 10^3 M_\odot$, modeled as a Schwarzschild black hole situated at the origin.  We construct the relaxed WD, which has a mass $M_{\rm wd} = 1.00 \, M_\odot$ and a radius of $R_{\rm wd} = 0.0082 \, R_\odot$. The center of mass of this relaxed WD is then placed at a distance of $5 r_t$ from the black hole, where $r_t$ is the tidal radius as mentioned earlier. We vary the impact parameter, denoted as $\beta = r_t/r_p$, ranging from $0.48$ to $1.00$, with each WD following a parabolic trajectory. Initially, the positions and velocities of all particles following the appropriate trajectories are calculated using the method explained in \cite{Banerjee2023}. 

In this work, one of our main goals will be to calculate the fallback rates while varying the impact parameter. When the WD approaches the black hole and has a relatively weak interaction at its closest point, some of the WD's material is disrupted. The extent of this disruption increases when the encounter is deeper, meaning that the pericenter distance is closer. We note that we are not concerned here with the circularization and disk formation processes. Therefore, after the interaction, once the WD moves away from the black hole and reaches a significant distance, we expand the accretion radius to $3 r_t \simeq 116 r_g$, where $r_g=G M_{\rm BH}/c^2$ represents the gravitational radius of the black hole. The black hole accretes any particle that enters this radius. Consequently, from our simulations, we obtain data on how the accreted mass varies over time. By taking the numerical derivative of this data, we determine the rate at which this accreted material falls onto the black hole. It is essential to note that the numerically calculated fallback rate differs from the true accretion rate, which represents the rate at which material crosses the black hole's horizon. To calculate the peak fallback rate directly from our simulations, we follow the methods outlined in \cite{Coughlin2015}, \cite{Golightly2019a}, \cite{Milesetal}, \cite{Garain2023}.

After the WD undergoes a weak interaction with the black hole and a portion of it is ejected, the remaining part of the star then contracts due to its own gravitational pull, forming a self-bound core. To identify these core particles, we employ an energy-based iterative method described by \cite{Guillochon2013}. The presence of the core leads to the formation of a high-density region, causing a significant reduction in the time step for the system's evolution. This poses a computational challenge for the system's evolution for longer duration. Therefore, when the surviving core moves a considerable distance away from the black hole and its properties are saturated, we substitute the core particles with a sink particle. The detailed procedure for introducing the sink particle is based on the methodology detailed in \cite{Garain2023}. We checked by changing the sink placement time to different values and observed no noticeable change in the fallback rate.

We employ $5\times 10^5$ particles to simulate the relaxed WD when the impact parameter falls between $0.60$ and $1.00$. However, during weak encounters, $\beta = 0.48, 0.50, 0.55$, only a small amount of material is disrupted. Hence, the debris contains very few particles. In such cases, we model the initial WD model using $10^6$ particles.

\section{Results}
\label{sec3}

In this section, we begin by determining the range of impact parameters at which partial disruption occurs when considering the EOS of the WD. In this context, it is well
known that the critical impact parameter ($\beta_c$) which distinguishes full disruption from partial disruption, depends strongly on the stellar structure and the star's 
rotation as well \citep{Guillochon2013, Mainetti2017, Golightly2019b}.

In our study of the tidal disruption of WDs by an IMBH, we employ a zero-temperature EOS to relate pressure and density, an improvement over the commonly used polytropic EOS in the literature. To determine $\beta_c$, we vary the impact parameter from $0.48$ to $1.00$. For each simulation, we calculate the mass loss, $\Delta M = M_{\rm wd} - M_{\rm core}$, when the core properties become saturated -- before incorporating the sink particle. 

\begin{figure}[h!]
\epsscale{0.8}
	\plottwo{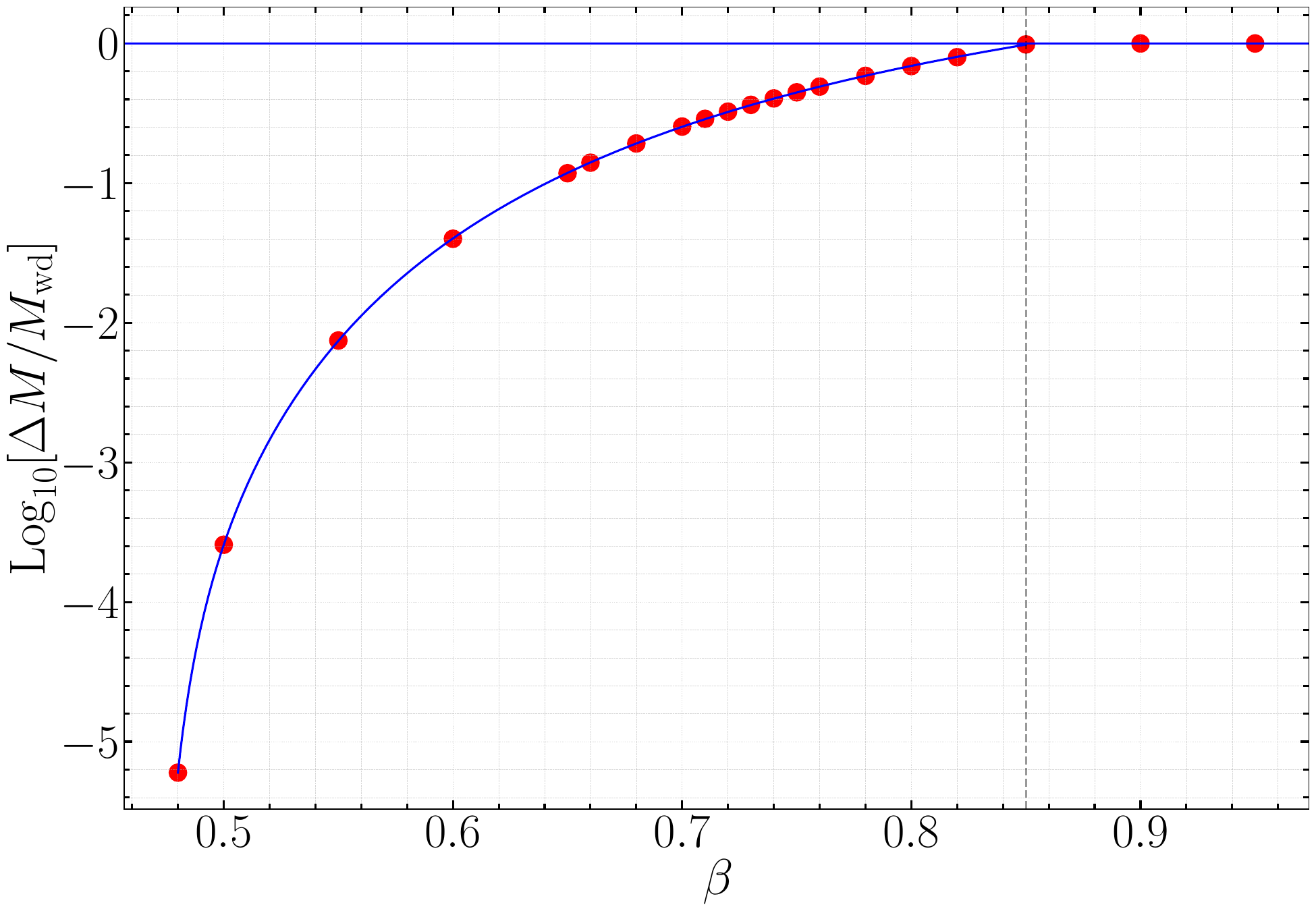}{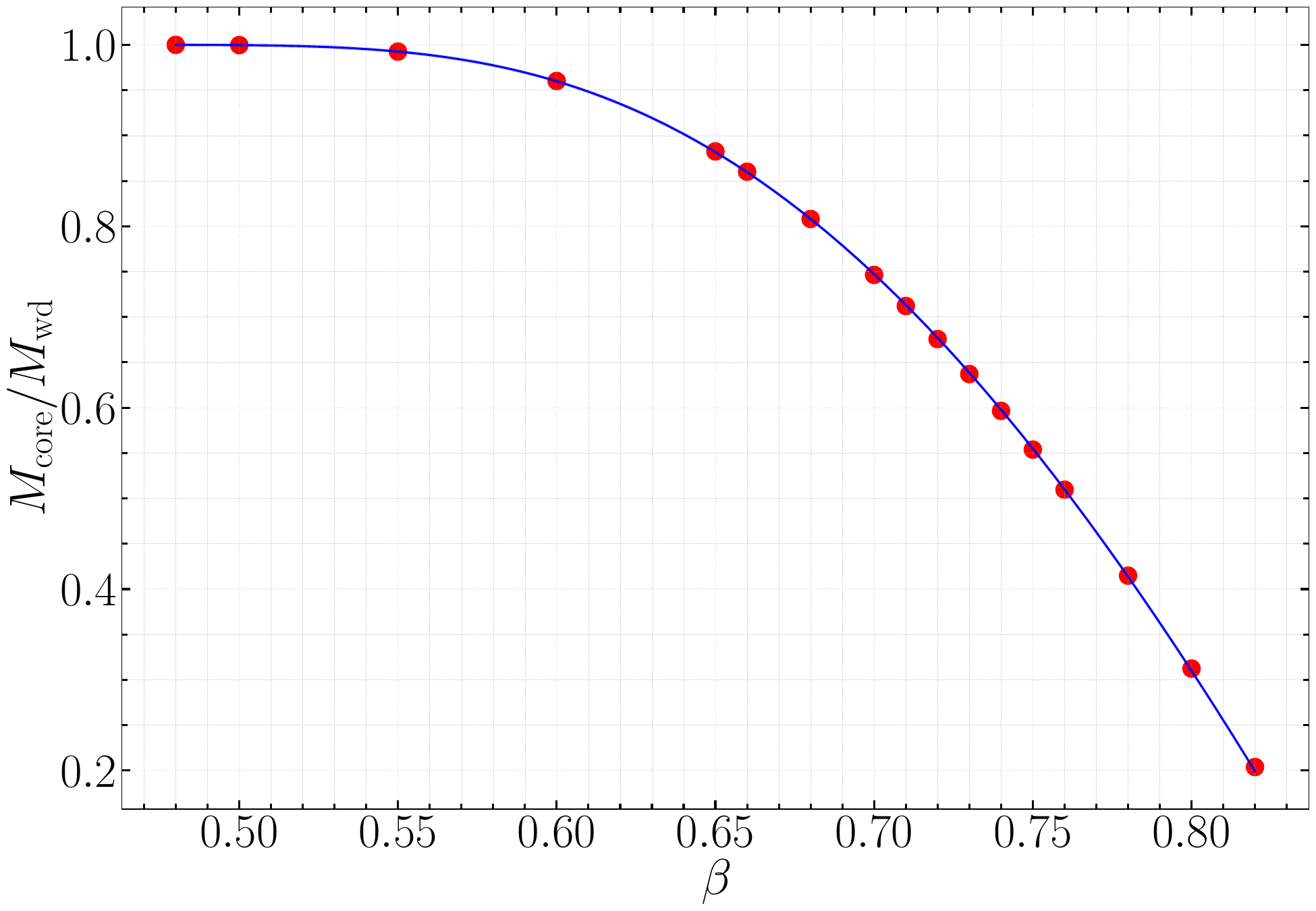}
	\caption{\textbf{Left panel:} Mass loss, $\Delta M$, relative to the initial WD mass, $M_{\rm wd}= 1M_\odot$, is plotted against $\beta$. A higher $\beta$ corresponds to a stronger tidal interaction, resulting in increased mass loss. The vertical black dashed line represents $\beta_c = 0.85$. \textbf{Right panel:} Variation of the final self-bound mass fraction, $M_{\rm core}/M_{\rm wd}$, with respect to $\beta$ is presented. No disruption occurs for $\beta < 0.48$.
	\label{fig.core}}
\end{figure}

In Figure \ref{fig.core}, the left panel shows the variation in $\Delta M$ relative to $M_{\rm wd}$ with changing $\beta$. Notably, for $\beta = 0.48$, the mass loss relative to the initial WD mass is merely $0.0006\%$, and any value of $\beta$ below $0.48$ results in no disruption. Thus $\beta = 0.48$ represents the lower limit for partial disruption in the context of this EOS. On the other hand, when $\beta < 0.85$, we observe a clear formation of a distinct core. Therefore, in our study, we determine the critical impact parameter to be $\beta_c = 0.85$, and any WD with an impact parameter falling in the range of $0.48 \leq \beta < 0.85$ is partially disrupted. In the right panel of Figure \ref{fig.core}, we present the variation of core mass relative to $M_{\rm wd}$ across different values of $\beta$.

We use a fitting function inspired by \cite{Guillochon2013} to model the variation of $\Delta M/M_{\rm wd}$ with respect to $\beta$:

\begin{eqnarray}
	\frac{\Delta M}{M_{\rm wd}} =
	\begin{cases}
		\exp{\Bigl[\frac{A + B \beta + C \beta^2}{1 - D \beta + E \beta^2 }\Bigr]} & 0.48\leq \beta < 0.85 \\
		1.0 & \beta \geq 0.85
	\end{cases}
	\label{Eq.deltaM}
\end{eqnarray}

and find the fitting parameters $A = 7.6678, B = -25.7050, C = 19.5925, D = 5.0697$, and $E = 6.2781$. 

However, it is important to note that the critical impact parameter can potentially deviate from this value in the presence of various factors, such as the rotation of the star or finite temperature corrections. However, our study provides a valuable starting point for modeling the WD with this specific EOS, leaving the exploration of these corrections to $\beta_c$ for future work. 

For comparison with \cite{Ryu2020c}, where they establish the fractional remnant mass ($M_{\rm rem}/M_\star$) as varying with the impact parameter according to a simple functional form of $\sim 1 - \beta^3$ for SMBH, we fit our $M_{\rm core}/M_{\rm wd}$ variation with $\beta$ using the form $1 - p\beta^q$. The obtained variation is well modeled by $\sim 1-3.1\beta^{6.8}$. Therefore, for an IMBH with a mass of $10^3 M_{\odot}$, one can represent the behavior as $\sim \beta^{6.8}$ when considering the variation of $\Delta M/M_{\rm wd}$ with $\beta$, while a more accurate fit is given by Equation (\ref{Eq.deltaM}).

As we determine the range of $\beta$ values where partial disruption occurs, we vary $\beta$ within the range of $0.55$ to $0.82$ and simulate partial tidal disruption. The disruptions that occur at $\beta = 0.48$ and $0.50$ are very weak, and the number of particles within the bound tail is extremely low, even with an initial $10^6$ particles. Therefore, the fallback rate curves exhibit significant noise, and we opt not to display these plots. We continue all simulations until approximately $97\%$ to $99\%$ of the bound material has accreted onto the black hole. Finally, numerically differentiating the accreted mass with time, we calculate the fallback rate for different $\beta$ values.

It is essential to note that when calculating the fallback rate, we do not rely on the snapshot method, which has several limitations. This method involves using the energy-period relationship to predict future orbits based on early time, assuming that the debris will follow these orbits while conserving its orbital energy throughout the trajectory. However, this assumption becomes inaccurate when a self-bound core is present, and there are scenarios where the absence of a core can also lead to inaccurate results, as shown by \cite{Coughlin2016a}, \cite{Coughlin2016b}. 

\begin{figure}[h!]
\epsscale{0.7}
	\plotone{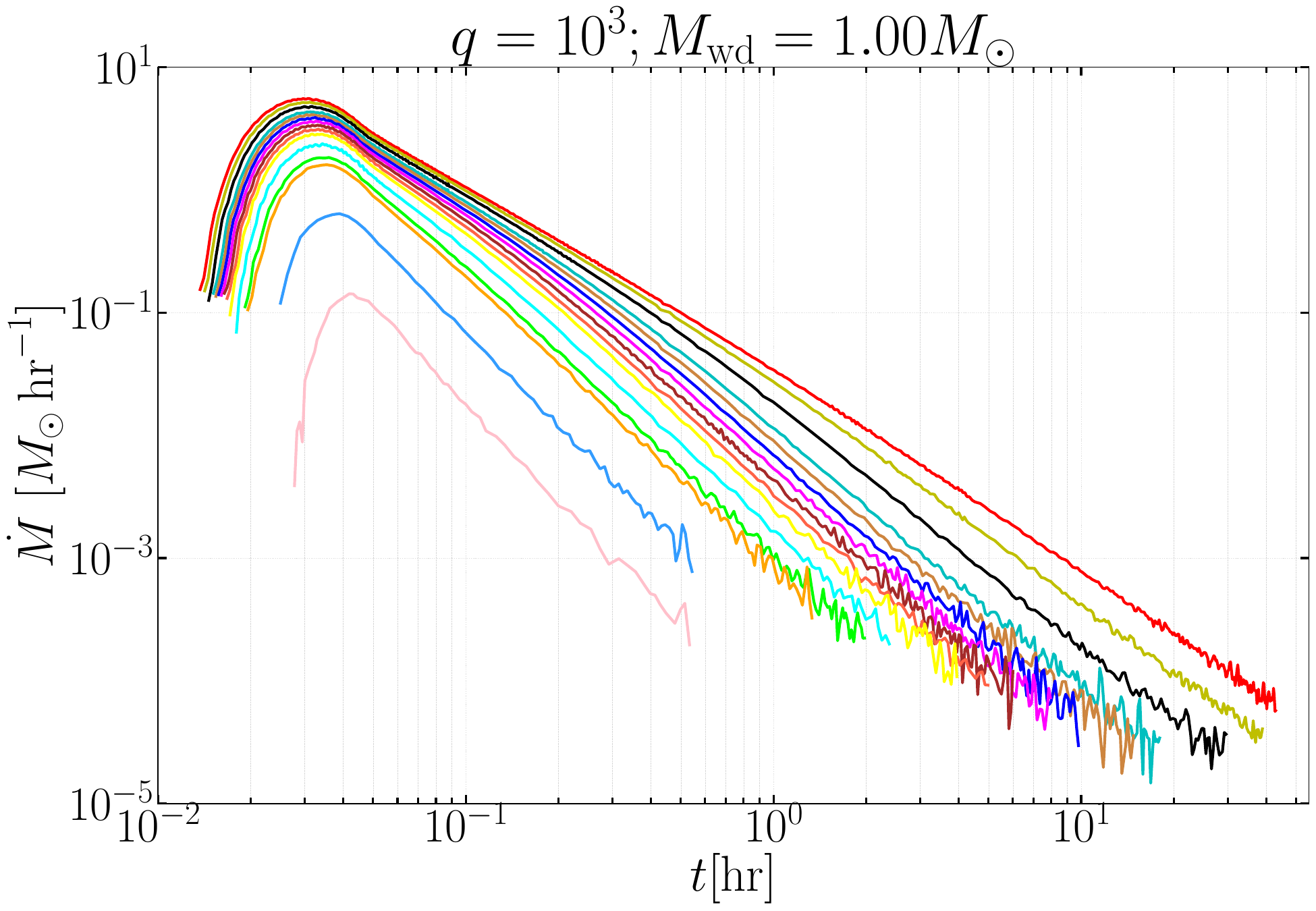}
	\plotone{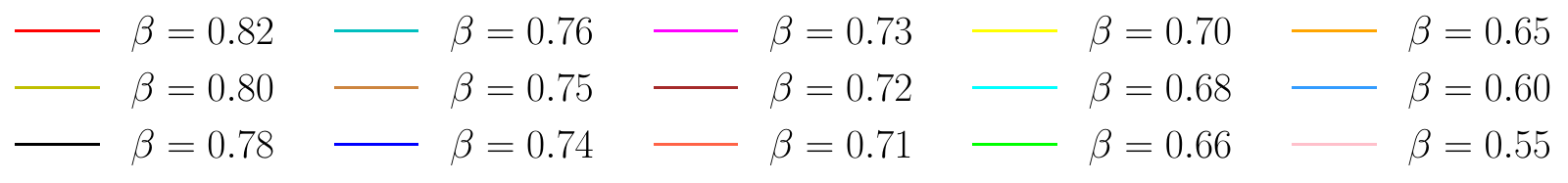}
	\caption{Fallback rates (in solar masses per hour) plotted as a function of time in hours for a $1 M_{\odot}$ WD due to an IMBH in 15 different simulations. These simulations cover a range of $\beta$ values mentioned in the legend from $0.55$ to $0.82$, resulting in partial disruptions. These curves clearly demonstrate that the fallback rates asymptote to different values.
	\label{fig.mdot}}
\end{figure}

In Figure \ref{fig.mdot}, the fallback rate is presented in Solar mass per hour as a function of time in hours for various $\beta$ values as mentioned in the legend. Note that the entire fallback of the bound debris occurs within a span of $\sim 100 \, {\rm hrs}$, which differs from the case of SMBHs, where the fallback process takes several years to complete.

\begin{figure}[h!]
\epsscale{0.8}
	\plottwo{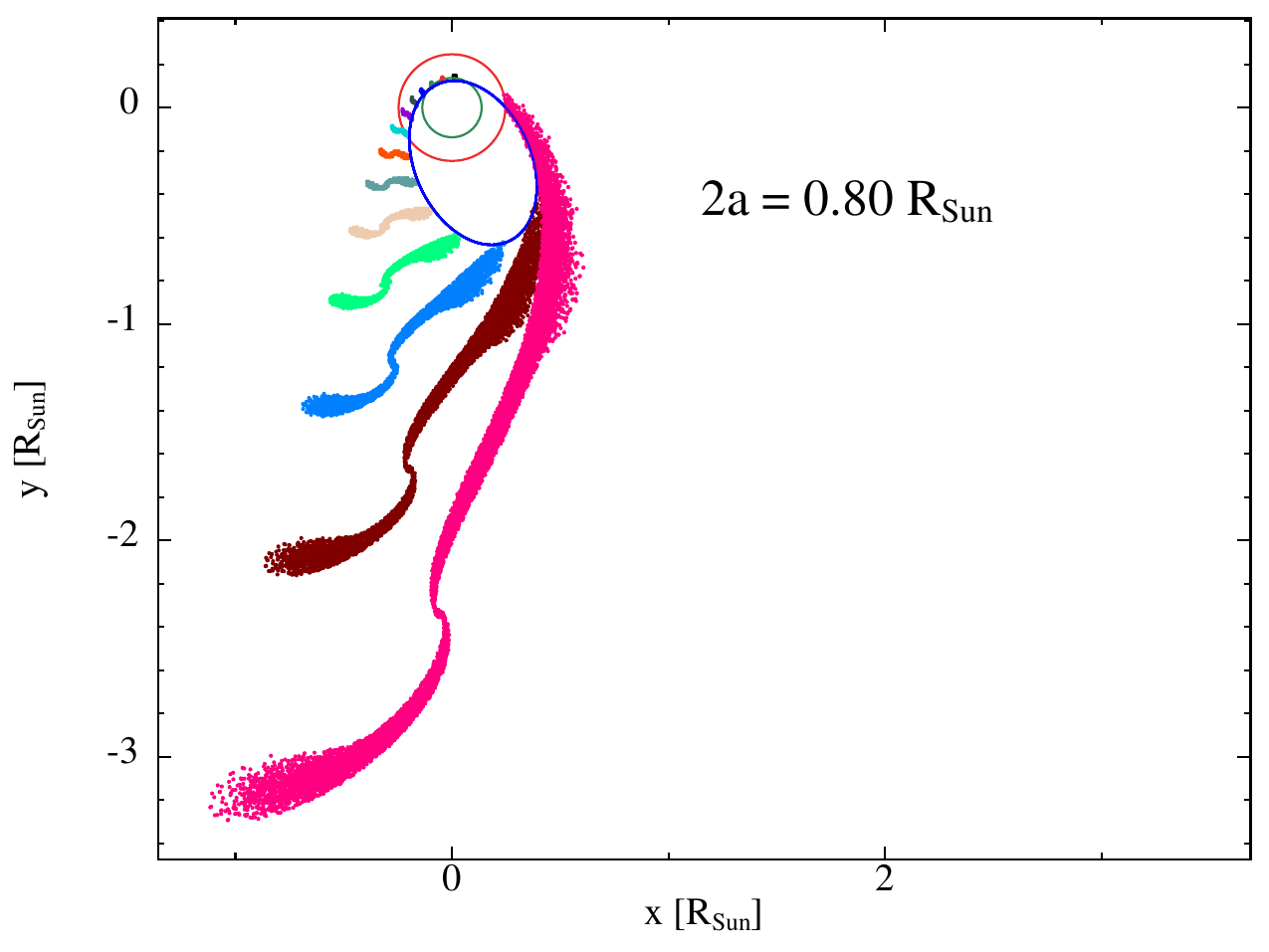}{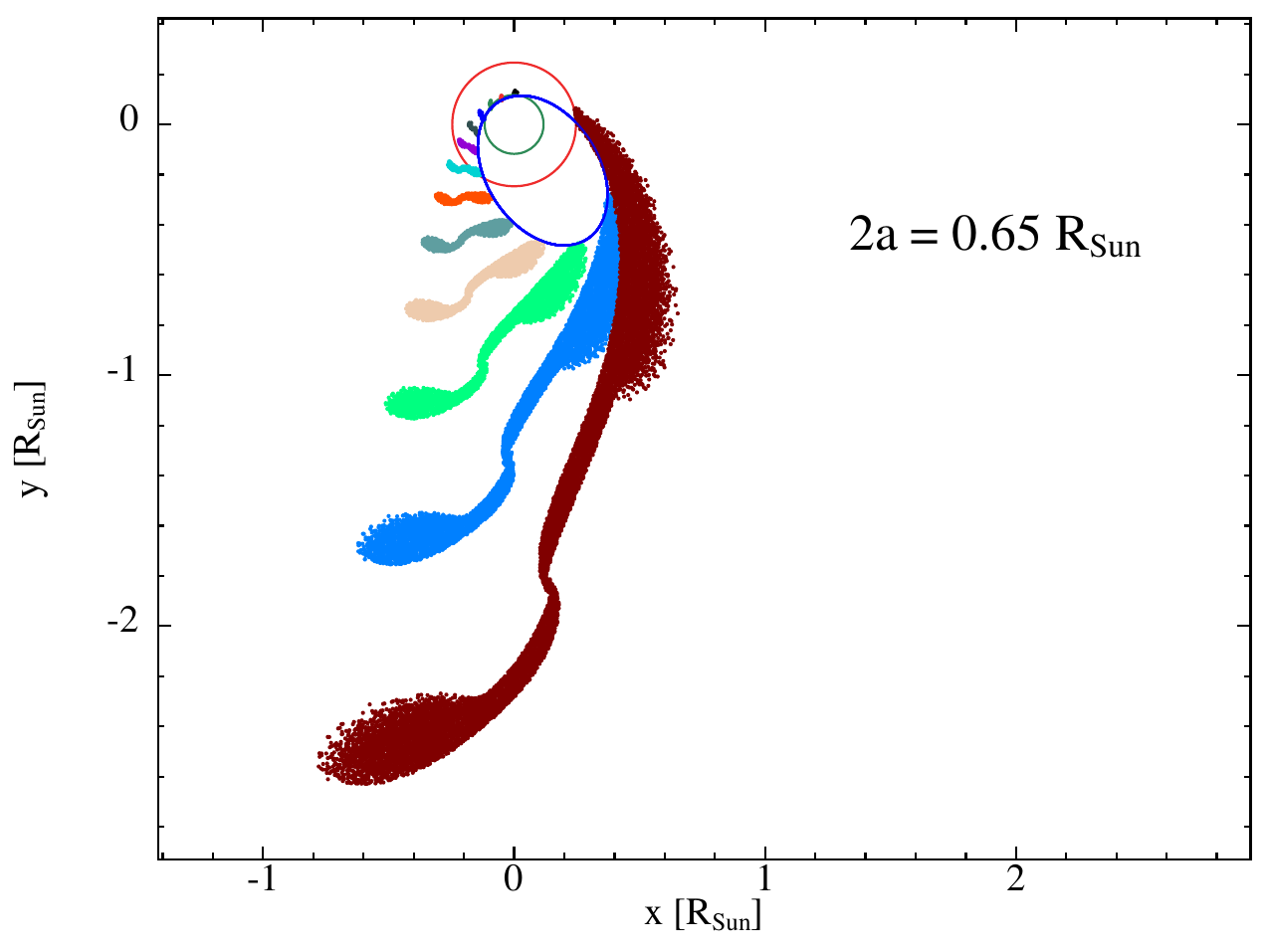}
	\plottwo{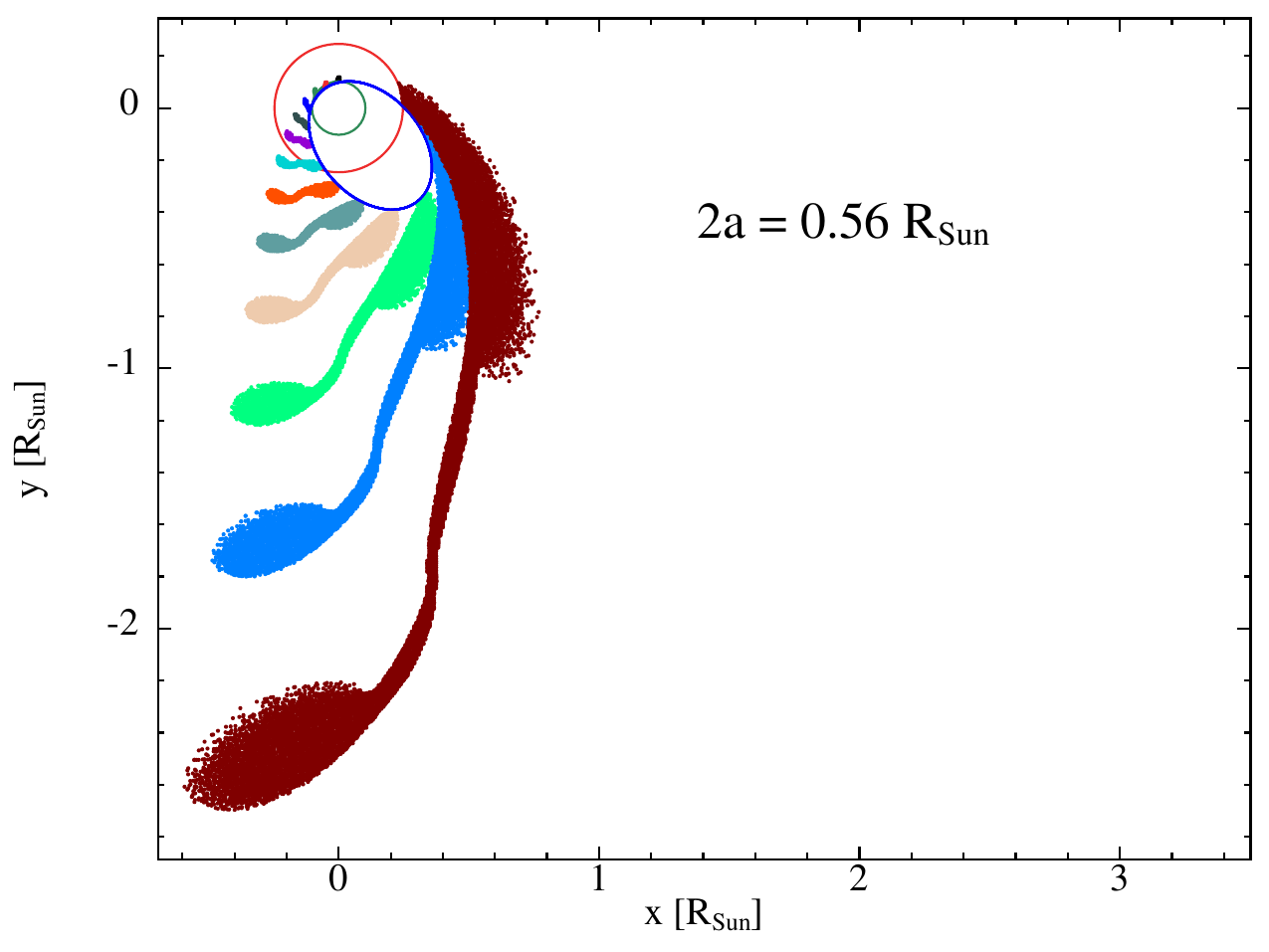}{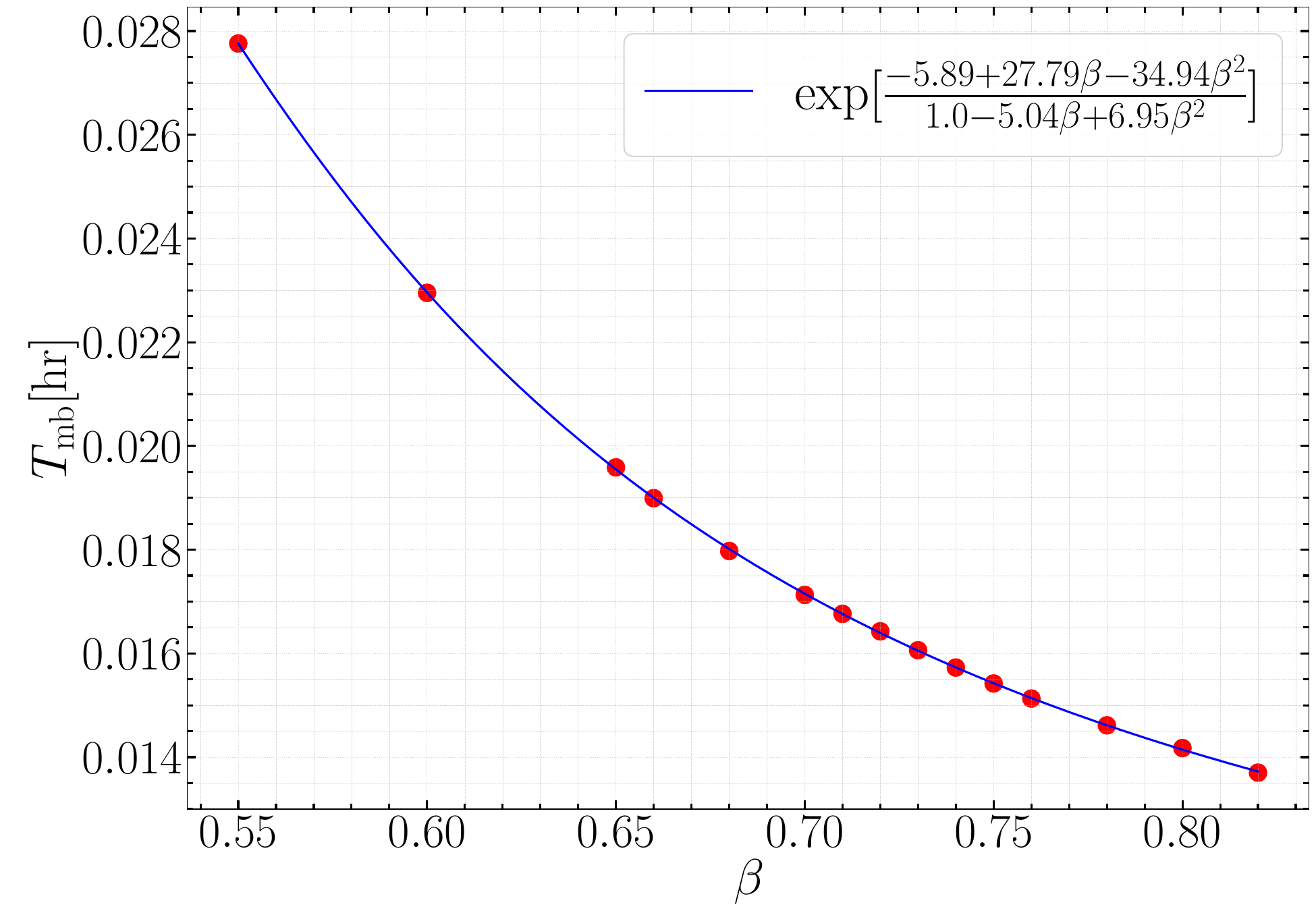}
	\caption{\textbf{Top Left panel:} Trajectory of the most bound debris is plotted for $\beta = 0.60$. \textbf{Top Right panel:} Trajectory of the most bound debris is plotted for $\beta = 0.70$. \textbf{Bottom Left panel:} Trajectory of the most bound debris is plotted for $\beta = 0.80$. Here, $a$ represents the semi-major axis of the orbits, and it is observed that as $\beta$ increases, the semi-major axis length decreases. These three figures are generated using SPLASH \citep{Price2007}. \textbf{Bottom Right panel:} The variation of the time at which the most bound debris falls onto the black hole ($T_{\rm mb}$) with $\beta$ is presented. This variation is modeled by the relation mentioned in the legend.
	\label{fig.Tmb}}
\end{figure}

The plot clearly shows that the time at which the most bound debris falls onto the black hole increases as $\beta$ decreases from $0.82$ to $0.55$. Following the tidal interaction, a particle in the maximally deformed region of star is positioned closer to the black hole with increasing $\beta$, as it interacts at a closer $r_p$ position. As a result, the most bound debris follows a trajectory that returns to the black hole more quickly as $\beta$ increases. We show the trajectories of the most bound debris for three different values of $\beta = 0.60, 0.70$, and $0.80$ in Figure \ref{fig.Tmb} Top Left, Top Right, and Bottom Left panel respectively. It is observed that with increasing $\beta$, the semi-major axis length of the most bound debris decreases, allowing it to reach the black hole more quickly. Figure \ref{fig.Tmb} Bottom Right panel presents the variation of the most bound debris fallback time ($T_{\rm mb}$) with respect to $\beta$.

\begin{figure}
\epsscale{0.8}
	\plottwo{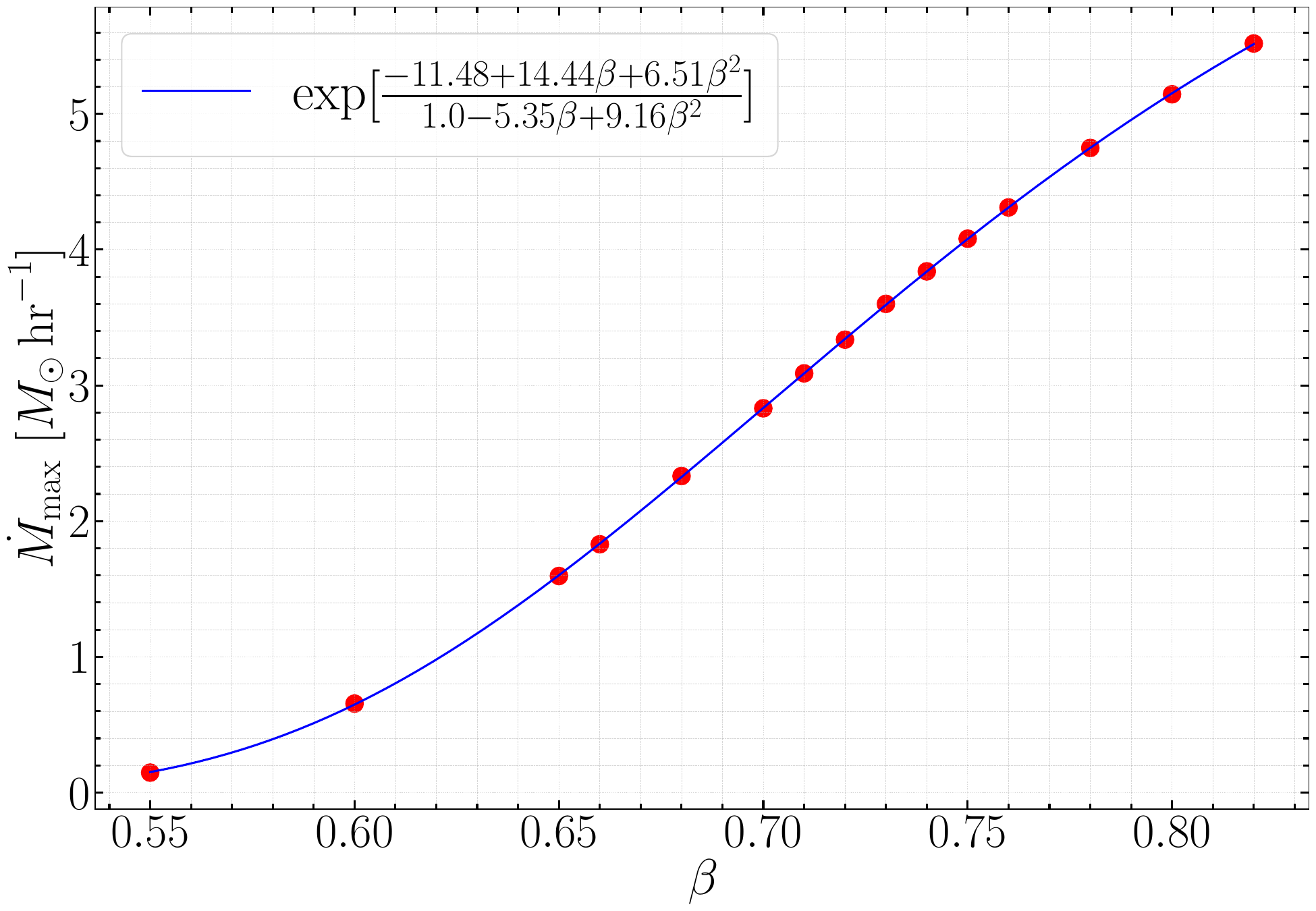}{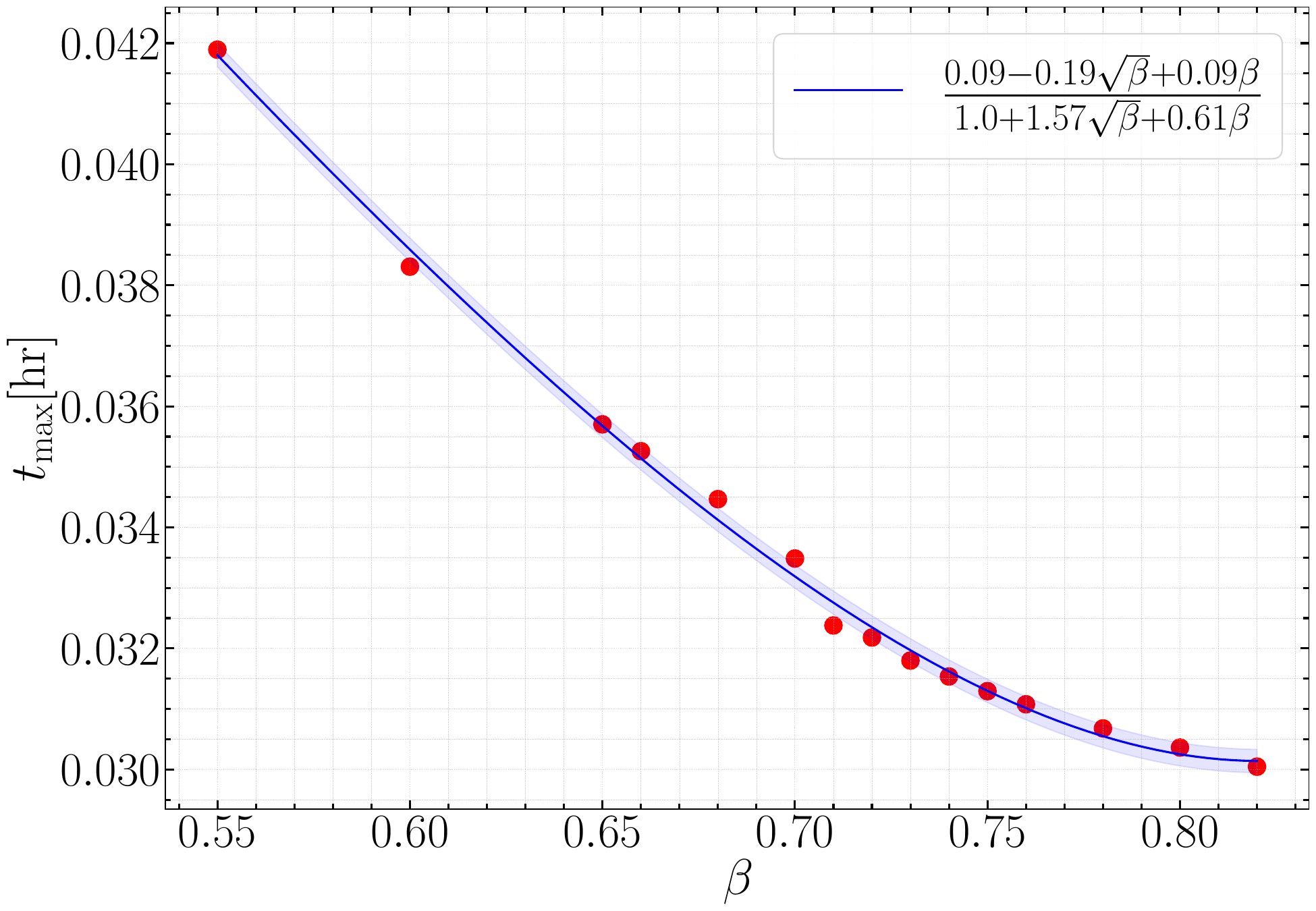}
	\caption{\textbf{Left panel:} Plot of $\dot{M}_{\rm max}$ as a function of $\beta$. \textbf{Right panel:} Plot of $t_{\rm max}$ as a function of $\beta$. As $\beta$ increases, $\dot{M}_{\rm max}$ significantly increases while $t_{\rm max}$ decreases slowly. The fitting functions are specified in the legend. The shaded blue region around the blue fitted curve represents the $1-\sigma$ deviation from the fit.
	\label{fig.mdotpeak}}
\end{figure}

We also depict the relationship between $\dot{M}_{\rm max}$ (the peak of the fallback rate) and $t_{\rm max}$ (the time at which the peak occurs) with respect to $\beta$ in Figure \ref{fig.mdotpeak} Left and Right panel respectively. It is evident that $\dot{M}_{\rm max}$ exhibits a strong dependence on $\beta$. Specifically, $\dot{M}_{\rm max, 0.82}$ is approximately $\sim 38$ times that of $\dot{M}_{\rm max, 0.55}$. However, the dependency of $t_{\rm max}$ on $\beta$ is relatively weak, with $t_{\rm max, 0.55}$ being only about $\sim 1.4$ times that of $t_{\rm max, 0.82}$.

For values of $\beta$ falling within the range of $0.65$ to $0.82$, the post-peak fallback curves exhibit an approximate scaling of $t^{-5/3}$, eventually steepening to $t^{n}$, where the exponent $n$ varies with $\beta$. Different values of $\beta$ lead to varying core masses and core-to-black hole mass ratios, defined as $\mu = M_{\rm core}/M_{\rm BH}$. In our study, $\mu$ ranges from $\sim 2\times10^{-4}$ to $10^{-3}$, whereas for SMBHs, $\mu$ is on the order of $\sim 10^{-6}$ to $10^{-7}$. This ratio is non-negligible for IMBHs and plays a significant role in determining the late-time slope, causing a deviation from the asymptotic scaling of $t^{-9/4}$ that applies to SMBHs.

\begin{figure}[h!]
\epsscale{0.7}
	\plotone{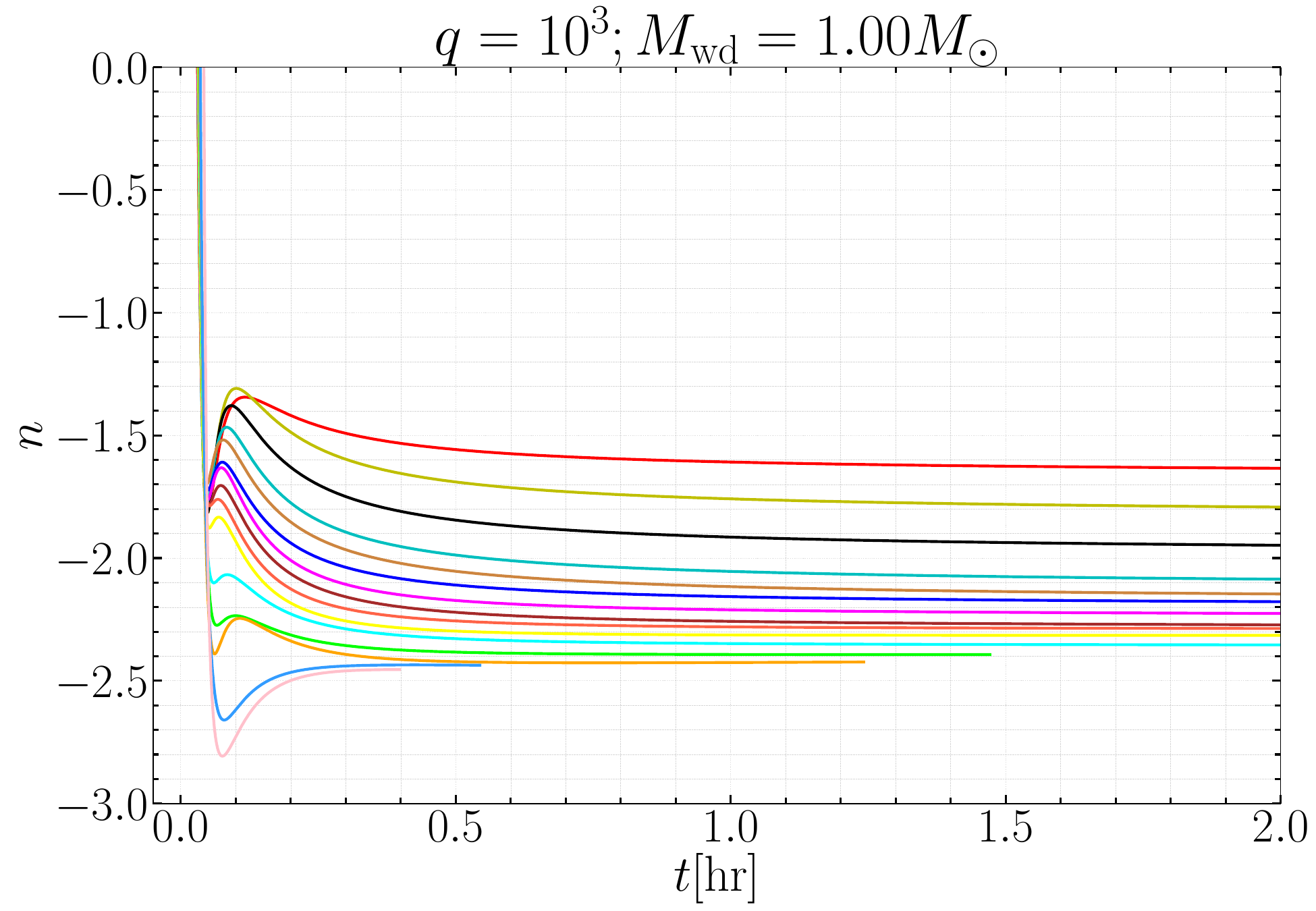}
	\plotone{legend.pdf}
	\caption{The instantaneous power-law index of the fallback rate, $n(t)$, is plotted against time in hours for various $\beta$ values mentioned in the legend. These plots are generated using the analytical fitting function described by Equation (\ref{Eq.modPade}) with $N_{\rm max} = 5$. It is evident that distinct $\beta$ values result in varying asymptotic power-law indices. 
	\label{fig.slopetime}}
\end{figure}

To accurately determine the late-time slope for different $\beta$ values, we follow the approach adopted by \cite{Nixon2021}. We fit the numerically obtain fallback rate with the modified Pad\'e approximant 

\begin{eqnarray}
	\dot{M}_{\rm fit} =	\frac{a\tilde{t}^m}{1+\frac{a}{b}\tilde{t}^{m-n_{\infty}}}\frac{1+\Sigma_{i=1}^{N_{\rm max}-1}c_i\tilde{t}^i+\tilde{t}^{N_{\rm max}}}{1+\tilde{t}^{N_{\rm max}}}
	\label{Eq.modPade}
\end{eqnarray}

Here, we introduce a normalized time variable, $\tilde{t} = t/t_{\rm max}$. This choice of analytical function, as suggested by \cite{Nixon2021}, is based on the observation that the initial rise before the peak and the late-time decay are effectively described by the functional form $a\tilde{t}^m/(1+\frac{a}{b}\tilde{t}^{m-n_{\infty}})$, where $\tilde{t}^m$ and $\tilde{t}^{n_{\infty}}$ represent power-law behaviors for the initial rise and late-time decay, respectively. The phase in between is accurately described by the ratio of polynomials, known as the Padé approximant. To determine the values of the constants $a, b, m, n_{\infty}, c_1, c_2, ..., c_{N_{\rm max}-1 }$, we minimize the chi-square defined as

\begin{eqnarray}
	\chi^2 = \sum_i\Bigg({\rm{ln}}\Big(\frac{\dot{M}_i}{\dot{M}_{\rm half max}}\Big) - {\rm{ln}}\Big(\frac{\dot{M}_{\rm fit}(\tilde{t}_i)}{\dot{M}_{\rm half max}}\Big)\Bigg)^2
\end{eqnarray}

Here, we define $\dot{M}_{\rm half max}$ as half of $\dot{M}_{\rm max}$, and $\dot{M}_i$ represents the numerically determined fallback rate at $\tilde{t}_i$. We tested the analytical fit using a range of terms, from $N_{\rm max} = 5$ to $10$. Our findings indicate that increasing the number of terms in the numerator of Equation (\ref{Eq.modPade}) does not significantly alter the characteristics of the curves. Moreover, the chi-square value does not decrease significantly when $N_{\rm max} > 5$. Therefore, we chose to use $N_{\rm max} = 5$ to model the fallback curves for different $\beta$ values. Once we obtain the analytical fits, we can determine the relationship between the power-law index and time using the simple relation $\dot{M} \propto t^{n}$, where $n(t) = d(\log{\dot{M}})/d(\log t)$. In Figure \ref{fig.slopetime}, we depict the variation in the power-law index over time in hours. It becomes evident from the figure that the late-time behavior of the power-law index ($n_{\infty}$) for the fallback rates with surviving cores does not asymptote to the well-established value of $-9/4$ for SMBHs. 

\begin{figure}[h!]
\epsscale{0.8}
	\plottwo{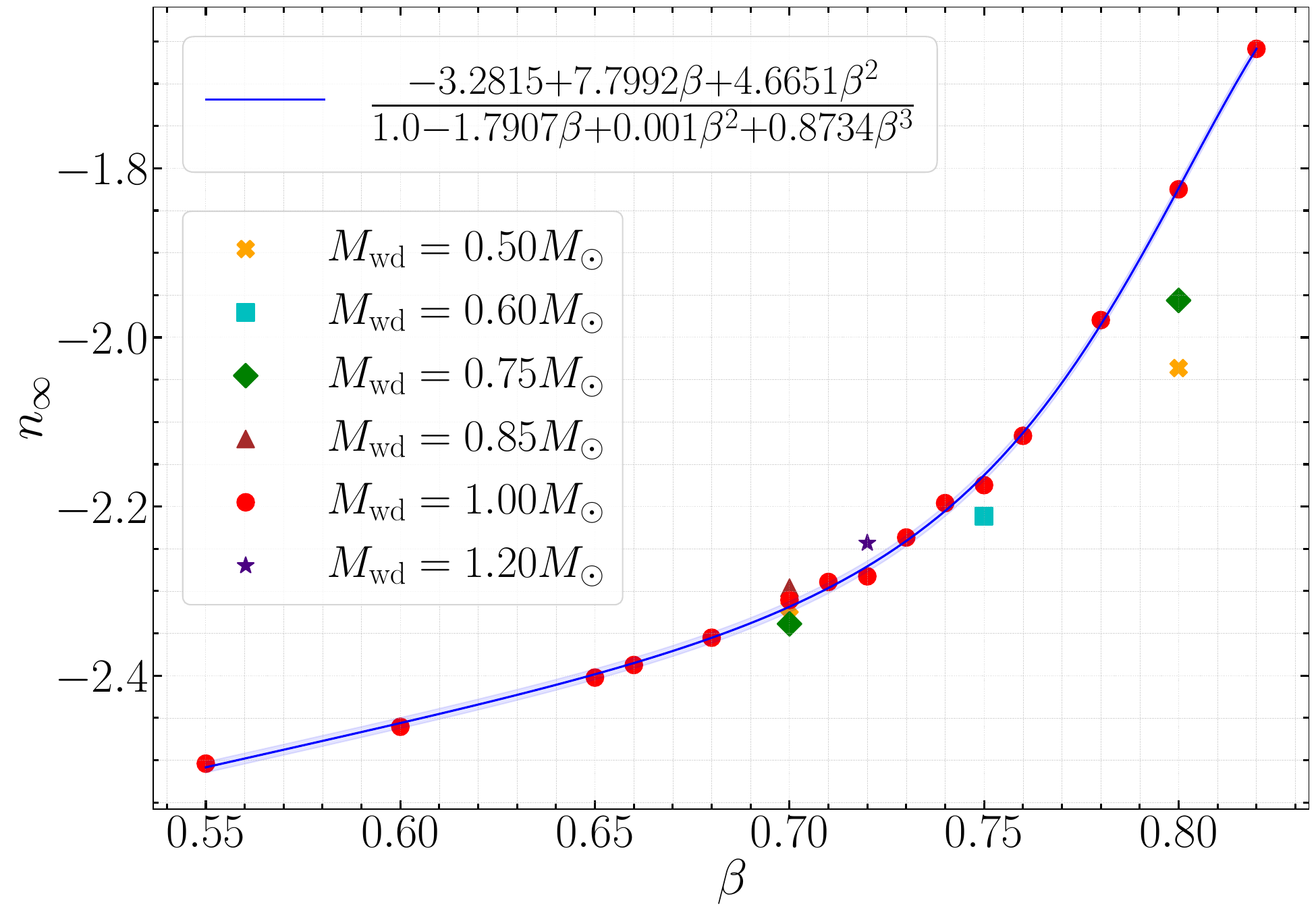}{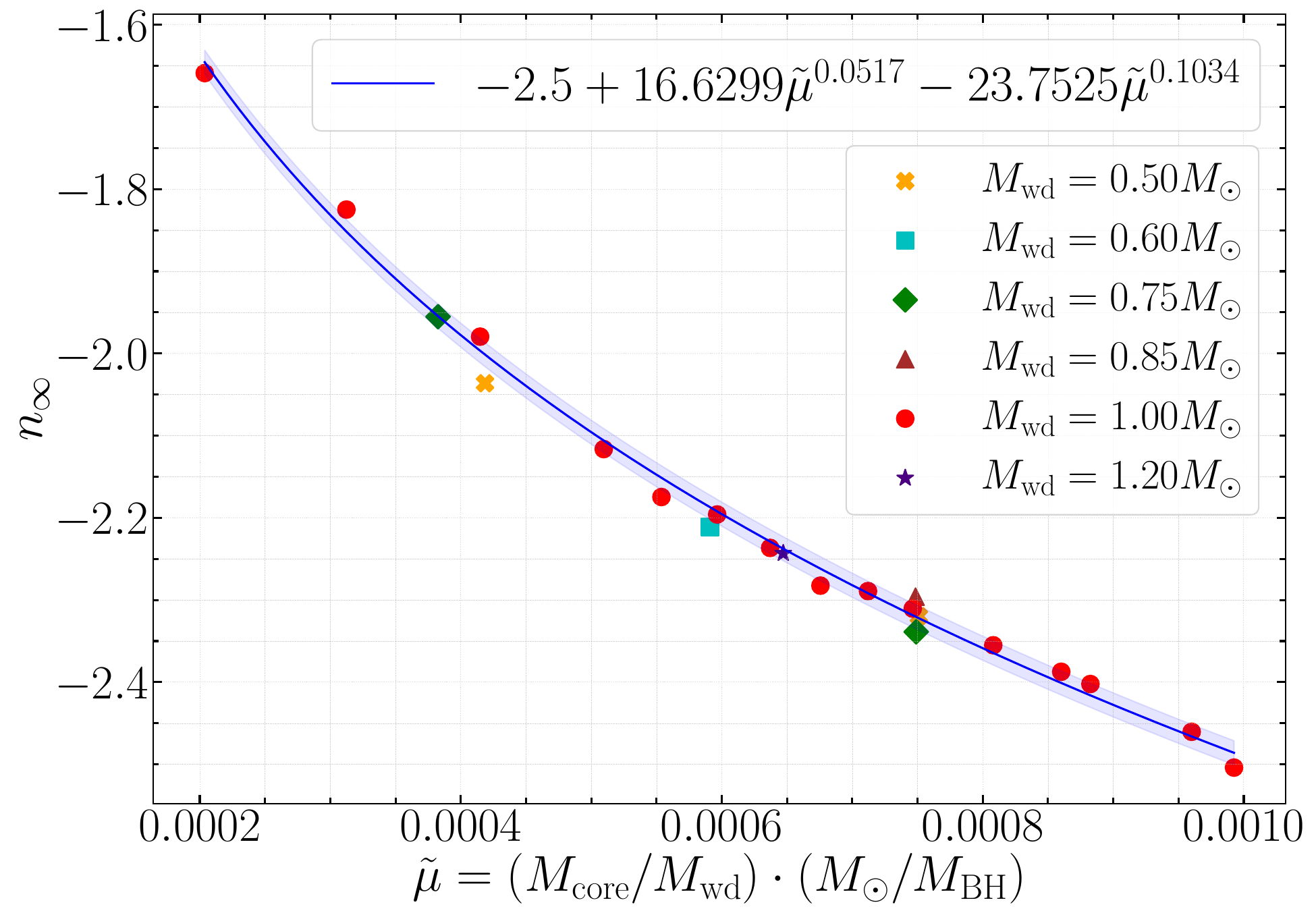}
	\caption{\textbf{Left panel:}  The late-time power-law index, $n_{\infty}$, is plotted as a function of $\beta$. \textbf{Right panel:} The late-time power-law index, $n_{\infty}$, is depicted as a function of $\tilde{\mu} = (M_{\rm core}/M_{\rm wd})\cdot(M_{\odot}/M_{\rm BH})$. The shaded blue region around the blue fitted curve represents the $1-\sigma$ deviation from the fit.
	\label{fig.slopefit}}
\end{figure}

We observe variations in $n_{\infty}$ for different $\beta$ values, ranging from $-1.66$ to $-2.50$ as $\beta$ decreases from $0.82$ to $0.55$. As core mass and $\mu$ both increase with decreasing $\beta$, the influence of the core on the falling material becomes more pronounced, leading to a deviation of $n_{\infty}$ from 
the value $-9/4$. 
In Figure \ref{fig.slopefit}, we present the variation of $n_{\infty}$ with $\beta$. We fit $n_{\infty}$ with respect to $\beta$, which is expressed as:

\begin{eqnarray}
	n_{\infty} = \frac{-3.2815 + 7.7992\beta + 4.6651\beta^2}{1 - 1.7907 \beta + 0.001 \beta^2 + 0.8734 \beta^3}.
	\label{Eq.nbeta}
\end{eqnarray}

We performed an additional set of $7$ simulations, systematically varying the initial WD mass and $\beta$ to assess the influence of changing stellar structure on $n_{\infty}$. The initial WD mass varies from $0.50 M_{\odot}$ to $1.20 M_{\odot}$, as indicated in the legend of Figure \ref{fig.slopefit}. We observe that $n_{\infty}$ exhibits notable variations with initial WD masses for higher $\beta$. 
For instance, with $\beta=0.70$ and initial WD masses of $0.50 M_{\odot}$ and $1.00 M_{\odot}$, we obtain $n_{\infty}$ values of $-2.32$ and $-2.31$, respectively. 
In contrast, for $\beta = 0.80$ with the same masses, $n_{\infty}$ becomes $-2.04$ and $-1.82$, respectively. This variation arises from the different amount of disruption observed for a fixed $\beta$ with different initial WD masses, and this discrepancy becomes more pronounced with increasing $\beta$. For an initial WD mass of $0.50 M_{\odot}$, the ratio of disrupted mass and the initial WD mass $\Delta M/M_{\rm wd}$ at $\beta = 0.70$ and $0.80$ is $0.25$ and $0.58$, respectively, while for $1.00 M_{\odot}$, these values are $0.25$ and $0.69$. Note that in the literature, typically the late time slope is studied as a function of the variable $\mu=
M_{\rm core}/M_{\rm BH}$. This variable however does not contain any information about the WD mass, since WDs with different masses and different
impact parameters can result in the same mass of the remnant core. Hence to make a more robust comparison of the late-time slopes for TDEs of
different WDs in the same BH background, we introduce a variable $\tilde{\mu} = (M_{\rm core}/M_{\rm wd})\cdot(M_{\odot}/M_{\rm BH})$. In Figure \ref{fig.slopefit}, we present the variation in $n_{\infty}$ with respect to $\tilde{\mu}$. We fit the relationship between $n_{\infty}$ and $\tilde{\mu}$ and provide the fitting formula as follows :

\begin{eqnarray}
	n_{\infty} =-2.5 + 16.6299\tilde{\mu}^{0.0517} - 23.7525\tilde{\mu}^{0.1034}.
	\label{Eq.nmu}
\end{eqnarray}

We studied partial disruption for 5 more WDs characterized by initial masses different from $1.00 M_{\odot}$ (see Figure \ref{fig.slopefit}). We see 
that the relationship between $n_{\infty}$ and $\tilde{\mu}$ closely matches the behaviour predicted by the fitting formula. It's noteworthy 
that all data points fall within the $3-\sigma$ range of the fitting formula, affirming the accuracy of the same.

\section{Discussion and Summary}
\label{sec4}

In this paper, we have studied partial TDEs, where an initial spherical WD in a parabolic orbit is partially disrupted by an IMBH. We have presented results of 22 numerical simulations.
In 15 of these, we considered a solar mass WD, and studied partial TDEs by varying the impact parameter $\beta$; and in the other 7 simulations, 
different WD masses were considered at particular values of $\beta$. We employed the zero temperature EOS directly in the SPH code to model the WDs, 
and finite temperature corrections have been neglected, as the central density of the initial WDs is $\sim 10^7 \, \text{g}\,\text{cm}^{-3}$. For solar mass
WDs, we determined the range of $\beta$ for which partial disruptions occur, using this EOS, and this gives $0.48 \leq \beta < 0.85$. 
We note that this range may be altered due to several factors, e.g., rotation of the 
initial WD which might modify our analysis, and we leave this issue for a separate study. 

We have calculated the fallback rates numerically for the partial TDEs using impact parameters ranging from $\beta = 0.55$ to $\beta = 0.82$, corresponding to 
pericenter positions from $70 r_g$ to $47 r_g$, where $r_g$ is the gravitational radius. According to \cite{Tejeda2017}, relativistic effects may be significant 
for $r_p \lesssim 10 r_g$. Therefore, in our study, we have safely neglected relativistic corrections. We have observed that with increasing $\beta$, the time of 
return of the most bound debris decreases. Additionally, we observe that the peak of the fallback rate strongly depends on $\beta$, increasing with higher values 
of $\beta$. However, the time at which the peak occurs only exhibits weak dependence on $\beta$ : as $\beta$ decreases, this time increases. 
Notably, we find here that the late-time power-law index does not asymptote to the usual $-9/4$ scaling, which is well known
for partial TDEs due to SMBHs. We have found that $n_{\infty}$ varies from $-5/3$ for full disruption to $-2.50$ for $\beta = 0.55$.  
This variation is due to the differing core masses for various $\beta$ values, and the ratio of the core mass to the black hole 
mass, although small, is not negligible, as is the case for SMBHs. This factor plays a significant role in determining $n_{\infty}$. 
Furthermore, we have found a robust fitting formula for the late time power law of the fallback rate (denoted by $n_{\infty}$) 
with the parameter $\tilde{\mu} = (M_{\rm core}/M_{\rm wd})\cdot(M_{\odot}/M_{\rm BH})$. 
Performing 7 simulations with initial WD mass other than $1 M_{\odot}$, we find that our formula accurately predicts the late time fall back rate,
independent of the WD mass, i.e., the obtained values fall within the $3-\sigma$ limit of the fitting formula. Changing the mass of the 
central black hole alters this formula, and issue that we have not delved into. 

We believe that the observational aspect of our study could be significant, since the fallback rate is expected to closely track the accretion rate. 
We have highlighted important differences of the physics of TDEs in the background of IMBHs as compared to SMBHs, and our results 
could serve as a robust criterion in the detection of IMBHs via TDEs, since deviations in the light curves from the expected $-9/4$ scaling 
could serve as a promising indicator for the presence of IMBHs.

\noindent
\begin{center}
{\bf Acknowledgements}\\
\end{center}
\noindent
We acknowledge the support and resources provided by PARAM Sanganak under the National Supercomputing Mission, Government of India, at the Indian Institute of Technology Kanpur. The work of DG is supported by grant number 09/092(1025)/2019-EMR-I from the Council of Scientific and Industrial Research (CSIR). The work of TS is supported
in part by the USV Chair Professor position at IIT Kanpur, India. 
\bigskip





\begin{thebibliography}{999}

\bibitem[Abbott et al.(2020)]{IMBH3} Abbott, R., Abbott, T.~D., Abraham, S., et al.\ 2020, \prl, 125, 101102. doi:10.1103/PhysRevLett.125.101102

\bibitem[Balsara(1995)]{Balsara1995} Balsara, D.~S.\ 1995, Journal of Computational Physics, 121, 357. doi:10.1016/S0021-9991(95)90221-X

\bibitem[Banerjee et al.(2023)]{Banerjee2023} Banerjee, P., Garain, D., Chowdhury, S., et al.\ 2023, \mnras, 522, 4332. doi:10.1093/mnras/stad1284

\bibitem[Bonnerot et al.(2016)]{Bonnerot2016} Bonnerot, C., Rossi, E.~M., Lodato, G., et al.\ 2016, \mnras, 455, 2253. doi:10.1093/mnras/stv2411

\bibitem[Brown et al.(2015)]{Brown} Brown, G.~C., Levan, A.~J., Stanway, E.~R., et al.\ 2015, \mnras, 452, 4297. doi:10.1093/mnras/stv1520

\bibitem[Carter \& Luminet(1982)]{CarterLumineta} Carter, B. \& Luminet, J.~P.\ 1982, \nat, 296, 211. doi:10.1038/296211a0

\bibitem[Carter \& Luminet(1983)]{CarterLuminetb} Carter, B. \& Luminet, J.-P.\ 1983, \aap, 121, 97

\bibitem[Chen \& Shen(2018)]{Chen2018} Chen, J.-H. \& Shen, R.-F.\ 2018, \apj, 867, 20. doi:10.3847/1538-4357/aadfda

\bibitem[Chen et al.(2023)]{Chen2023} Chen, J.-H. Shen, R.-F. \& Liu S-F., \apj, 947, 32. doi:10.3847/1538-4357/acbfb6

\bibitem[Chilingarian et al.(2018)]{IMBH1} Chilingarian, I.~V., Katkov, I.~Y., Zolotukhin, I.~Y., et al.\ 2018, \apj, 863, 1. doi:10.3847/1538-4357/aad184

\bibitem[Clerici \& Gomboc(2020)]{Clerici2020} Clerici, A. \& Gomboc, A.\ 2020, \aap, 642, A111. doi:10.1051/0004-6361/202037641
 
\bibitem[Coughlin \& Nixon(2015)]{Coughlin2015} Coughlin, E.~R. \& Nixon, C.\ 2015, \apjl, 808, L11. doi:10.1088/2041-8205/808/1/L11.

\bibitem[Coughlin et al.(2016a)]{Coughlin2016a} Coughlin, E.~R., Nixon, C., Begelman, M.~C., et al.\ 2016a, \mnras, 455, 3612. doi:10.1093/mnras/stv2511

\bibitem[Coughlin et al.(2016b)]{Coughlin2016b} Coughlin, E.~R., Nixon, C., Begelman, M.~C., et al.\ 2016b, \mnras, 459, 3089. doi:10.1093/mnras/stw770

\bibitem[Coughlin \& Nixon(2019)]{CoughlinNixon} Coughlin, E.~R. \& Nixon, C.~J.\ 2019, \apjl, 883, L17. doi:10.3847/2041-8213/ab412d

\bibitem[Cufari et al.(2022)]{Cufari2022} Cufari, M., Coughlin, E.~R., \& Nixon, C.~J.\ 2022, \apj, 924, 34. doi:10.3847/1538-4357/ac32be

\bibitem[Darbha et al.(2019)]{Darbha2019} Darbha, S., Coughlin, E.~R., Kasen, D., et al.\ 2019, \mnras, 488, 5267. doi:10.1093/mnras/stz1923

\bibitem[Evans \& Kochanek(1989)]{Evans1989} Evans, C.~R. \& Kochanek, C.~S.\ 1989, \apjl, 346, L13. doi:10.1086/185567

\bibitem[Frank \& Rees(1976)]{FrankRees} Frank, J. \& Rees, M.~J.\ 1976, \mnras, 176, 633. doi:10.1093/mnras/176.3.633

\bibitem[Gafton et al.(2015)]{Gafton2015} Gafton, E., Tejeda, E., Guillochon, J., et al.\ 2015, \mnras, 449, 771. doi:10.1093/mnras/stv350

\bibitem[Gafton \& Rosswog(2019)]{Gafton2019} Gafton, E. \& Rosswog, S.\ 2019, \mnras, 487, 4790. doi:10.1093/mnras/stz1530

\bibitem[Garain et al.(2023)]{Garain2023} Garain, D., Banerjee, P., Chowdhury, S., et al.\ 2023, arXiv:2307.03142. doi:10.48550/arXiv.2307.03142

\bibitem[Golightly et al.(2019a)]{Golightly2019a} Golightly, E.~C.~A., Nixon, C.~J., \& Coughlin, E.~R.\ 2019a, \apjl, 882, L26. doi:10.3847/2041-8213/ab380d

\bibitem[Golightly et al.(2019b)]{Golightly2019b} Golightly, E.~C.~A., Coughlin, E.~R., \& Nixon, C.~J.\ 2019b, \apj, 872, 163. doi:10.3847/1538-4357/aafd2f

\bibitem[Greene et al.(2020)]{Green} Greene, J.~E., Strader, J., \& Ho, L.~C.\ 2020, \araa, 58, 257. doi:10.1146/annurev-astro-032620-021835

\bibitem[Guillochon \& Ramirez-Ruiz(2013)]{Guillochon2013} Guillochon, J. \& Ramirez-Ruiz, E.\ 2013, \apj, 767, 25. doi:10.1088/0004-637X/767/1/25

\bibitem[Guillochon et al.(2014)]{Guillochon2014} Guillochon, J., Manukian, H., \& Ramirez-Ruiz, E.\ 2014, \apj, 783, 23. doi:10.1088/0004-637X/783/1/23

\bibitem[Hayasaki et al.(2013)]{Hayasaki2013} Hayasaki, K., Stone, N., \& Loeb, A.\ 2013, \mnras, 434, 909. doi:10.1093/mnras/stt871

\bibitem[Hayasaki et al.(2016)]{Hayasaki2016} Hayasaki, K., Stone, N., \& Loeb, A.\ 2016, \mnras, 461, 3760. doi:10.1093/mnras/stw1387

\bibitem[Hills(1975)]{Hills} Hills, J.~G.\ 1975, \nat, 254, 295. doi:10.1038/254295a0

\bibitem[Holoien et al.(2019)]{Holoien} Holoien, T.~W.-S., Vallely, P.~J., Auchettl, K., et al.\ 2019, \apj, 883, 111. doi:10.3847/1538-4357/ab3c66

\bibitem[Jonker et al.(2022)] {Jonker} Jonker, P.~G, Arcavi, I., Phinney, E.~S, et al.\ 2022, The Tidal Disruption of Stars by Massive Black Holes (Springer) 

\bibitem[Kagaya et al.(2019)]{Kagaya2019} Kagaya, K., Yoshida, S., \& Tanikawa, A.\ 2019, arXiv:1901.05644. doi:10.48550/arXiv.1901.05644

\bibitem[K{\i}z{\i}ltan et al.(2017)]{IMBH0} K{\i}z{\i}ltan, B., Baumgardt, H., \& Loeb, A.\ 2017, \nat, 542, 203. doi:10.1038/nature21361

\bibitem[Kochanek(1994)]{Kochanek1994} Kochanek, C.~S.\ 1994, \apj, 422, 508. doi:10.1086/173745

\bibitem[Lacy et al.(1982)]{Lacy1} Lacy, J.~H., Townes, C.~H., \& Hollenbach, D.~J.\ 1982, \apj, 262, 120. doi:10.1086/160402

\bibitem[Law-Smith et al.(2017)]{Law2017} Law-Smith, J., MacLeod, M., Guillochon, J., et al.\ 2017, \apj, 841, 132. doi:10.3847/1538-4357/aa6ffb

\bibitem[Law-Smith et al.(2019)]{Law2019} Law-Smith, J., Guillochon, J., \& Ramirez-Ruiz, E.\ 2019, \apjl, 882, L25. doi:10.3847/2041-8213/ab379a

\bibitem[Law-Smith et al.(2020)]{Law2020} Law-Smith, J.~A.~P., Coulter, D.~A., Guillochon, J., et al.\ 2020, \apj, 905, 141. doi:10.3847/1538-4357/abc489

\bibitem[Lin et al.(2020)]{IMBH4} Lin, D., Strader, J., Romanowsky, A.~J., et al.\ 2020, \apjl, 892, L25. doi:10.3847/2041-8213/ab745b

\bibitem[Liptai et al.(2019)]{Liptai2019} Liptai, D., Price, D.~J., Mandel, I., et al.\ 2019, arXiv:1910.10154. doi:10.48550/arXiv.1910.10154
 
\bibitem[Lodato et al.(2009)]{Lodato} Lodato, G., King, A.~R., \& Pringle, J.~E.\ 2009, \mnras, 392, 332. doi:10.1111/j.1365-2966.2008.14049.x

\bibitem[MacLeod et al.(2013)]{MacLeod2013} MacLeod, M., Ramirez-Ruiz, E., Grady, S., et al.\ 2013, \apj, 777, 133. doi:10.1088/0004-637X/777/2/133

\bibitem[Mainetti et al.(2017)]{Mainetti2017} Mainetti, D., Lupi, A., Campana, S., et al.\ 2017, \aap, 600, A124. doi:10.1051/0004-6361/201630092

\bibitem[Manukian et al.(2013)]{Manukian2013} Manukian, H., Guillochon, J., Ramirez-Ruiz, E., et al.\ 2013, \apjl, 771, L28. doi:10.1088/2041-8205/771/2/L28

\bibitem[Miles et al.(2020)]{Milesetal} Miles, P.~R., Coughlin, E.~R., \& Nixon, C.~J.\ 2020, \apj, 899, 36. doi:10.3847/1538-4357/ab9c9f

\bibitem[Nixon et al.(2021)]{Nixon2021} Nixon, C.~J., Coughlin, E.~R., \& Miles, P.~R.\ 2021, \apj, 922, 168. doi:10.3847/1538-4357/ac1bb8

\bibitem[Park \& Hayasaki(2020)]{Park2020} Park, G. \& Hayasaki, K.\ 2020, \apj, 900, 3. doi:10.3847/1538-4357/ab9ebb

\bibitem[Phinney(1989)]{Phinney} Phinney, E.~S.\ 1989, The Center of the Galaxy, 136, 543

\bibitem[Price(2007)]{Price2007} Price, D.~J.\ 2007, \pasa, 24, 159. doi:10.1071/AS07022

\bibitem[Rees(1988)]{Rees} Rees, M.~J.\ 1988, \nat, 333, 523. doi:10.1038/333523a0 
.
\bibitem[Ryu et al.(2020a)]{Ryu2020a} Ryu, T., Krolik, J., Piran, T., et al.\ 2020a, \apj, 904, 98. doi:10.3847/1538-4357/abb3cf

\bibitem[Ryu et al.(2020b)]{Ryu2020b} Ryu, T., Krolik, J., Piran, T., et al.\ 2020b, \apj, 904, 100. doi:10.3847/1538-4357/abb3ce

\bibitem[Ryu et al.(2020c)]{Ryu2020c} Ryu, T., Krolik, J., Piran, T., et al.\ 2020c, \apj, 904, 100. doi:10.3847/1538-4357/abb3ce

\bibitem[Sacchi \& Lodato(2019)]{Sacchi2019} Sacchi, A. \& Lodato, G.\ 2019, \mnras, 486, 1833. doi:10.1093/mnras/stz981

\bibitem[Takekawa et al.(2019)]{IMBH2} Takekawa, S., Oka, T., Iwata, Y., et al.\ 2019, \apjl, 871, L1. doi:10.3847/2041-8213/aafb07

\bibitem[Tejeda et al.(2017)]{Tejeda2017} Tejeda, E., Gafton, E., Rosswog, S., et al.\ 2017, \mnras, 469, 4483. doi:10.1093/mnras/stx1089

\bibitem[Tejeda \& Rosswog(2013)]{Rosswog_TR} Tejeda, E. \& Rosswog, S.\ 2013, \mnras, 433, 1930. doi:10.1093/mnras/stt853

\bibitem[Open TDE Catalog (2023)]{OpenTDE} 
The Open TDE Catalog, 2023, available at https://tde.space/

\bibitem[Ulmer(1999)]{Ulmer1999} Ulmer, A.\ 1999, \apj, 514, 180. doi:10.1086/306909.

\bibitem[Volonteri(2012)]{Volonteri} Volonteri, M.\ 2012, Science, 337, 544. doi:10.1126/science.1220843

\bibitem[Wang et al.(2021)]{Wang2021} Wang, Y.-H., Perna, R., \& Armitage, P.~J.\ 2021, \mnras, 503, 6005. doi:10.1093/mnras/stab802.





\end{thebibliography}
\end{document}